\documentclass[aps,pra,twocolumn,superscriptaddress,longbibliography,amsmath,amssymb,amsfonts,citeautoscript,10pt]{revtex4-2}
\usepackage{graphicx}
\usepackage{bm}
\usepackage{color}
\usepackage{epstopdf}
\usepackage{amsmath}
\usepackage{amssymb}
\usepackage{ulem}
\usepackage[urlcolor=blue,colorlinks=true,citecolor=blue,linkcolor=blue,pdfstartview={FitH},bookmarks=false]{hyperref}
\usepackage{ulem}

\newcommand{\e}{\textrm{e}}

\newcommand{\Imag}{\text{Im}}

\newcommand{\be}{\begin{equation}}
\newcommand{\ee}{\end{equation}}
\newcommand{\bea}{\begin{eqnarray}}
\newcommand{\eea}{\end{eqnarray}}

\newcommand{\en}{\varepsilon}

\newcommand{\s}{\sigma}

\newcommand{\up}{\uparrow}
\newcommand{\down}{\downarrow}

\renewcommand{\Re}{\mathrm{Re}}

\newcommand{\nn}{\nonumber}
\newcommand{\bk}{\mathbf{k}}

\usepackage{ulem}
\usepackage{verbatim} 

\begin{document}

\title{Properties of multiterminal superconducting nanostructure
with double quantum dot}

\author{G.\ G\'{o}rski}
\email{ggorski@ur.edu.pl}
\affiliation{Institute of Physics, Faculty of Exact and Technical Sciences, University of Rzesz\'{o}w, 35-310 Rzesz\'{o}w, Poland}
\author{K.\ Kucab}
\affiliation{Institute of Physics, Faculty of Exact and Technical Sciences, University of Rzesz\'{o}w, 35-310 Rzesz\'{o}w, Poland}
\author{T.\ Doma\'nski}
\email{doman@kft.umcs.lublin.pl}
\affiliation{Institute of Physics, M.\ Curie-Sk\l{}odowska University, 20-031 Lublin, Poland}

\date{\today}

\begin{abstract}
We study the charge transport and thermoelectric properties of the junction, comprising double quantum dot embedded in T-shaped geometry on the interface between two normal/ferromagnetic electrodes and  superconducting lead. We show that the interdot coupling plays major role in controlling the local and nonlocal transport properties of this setup. For the weak interdot coupling limit, we obtain the interferometric (Fano-type) lineshapes imprinted in the quasiparticle spectra, conductances and Seebeck coefficients. In contrast, for the strong interdot coupling, we predict that the local and nonlocal transport coefficients are primarily dependent on the molecular Andreev bound states induced by superconducting proximity effect, simultaneously in both quantum dots. 
\end{abstract}

\maketitle

\section{Introduction}
\label{sec:introduction}

Charge transport through superconducting heterostructures, comprising the quantum dots (QDs), is recently intensively explored due to perspectives of possible application in nanoelectronics, spintronics, metrology, and quantum information processing \cite{Krantz-2019}. Various configurations of QDs coupled either to conventional \cite{Benito-2020,Mazur-2024} or topological superconductors \cite{DasSarma-2016,Aguado-2020} are considered, offering realization of brand new technological devices. 

Transport properties of hybrid structures where QDs are between the superconducting (S) and normal (N) or ferromagnetic (F)  electrodes are essentially affected by the bound states \cite{Bauer-2007,Martin-Rodero-2011}, enabling the subgap charge transfer via electron-to-hole (Andreev) scattering \cite{Eichler-2007,Deacon-2010,Pillet-2010,Dirks-2011}. Such in-gap states originate from the superconducting proximity effect. Competition with the on-dot Coulomb repulsion, however, can lead under specific conditions to the single occupancy of QD, allowing for the Kondo state to emerge \cite{Bauer-2007,Zitko-2015,Domanski-2016}. Signatures of these  Andreev and Kondo effects have been observed in various nanostructures  \cite{Pillet-2013,Kim-2013,He-2020}. By varying the energy level or hybridization to external leads the ground state of QD can change from the single occupied to the BCS-type configuration what is manifested by a crossing of the in-gap states \cite{Bauer-2007}. At such parity crossing the low-temperature Andreev conductance reaches its optimal value  $4e^2/h$  \cite{Yamada-2010,Yamada-2011,Domanski-2016,Gorski-2018,Hashimoto-2024}. Thermal excitations can further activate the quasiparticle excitations from outside the pairing gap, giving rise to the Seebeck effect \cite{Hwang-2016,Hwang-2016B,Verma-2024}.

\begin{figure}
\centering
\includegraphics[width=0.7\linewidth]{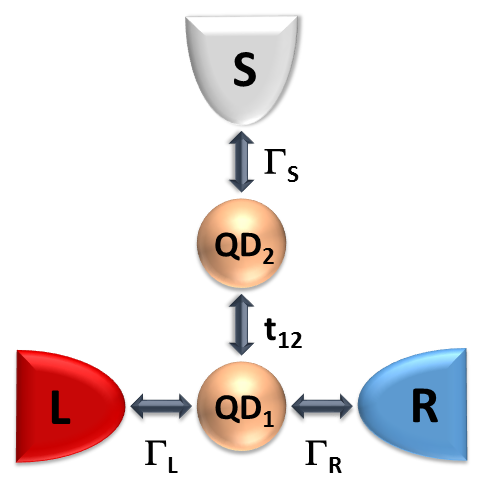}
\caption{Sketch of two quantum dots (QD$_1$ and QD$_2$) on interface of three-terminal junction. QD$_1$ is embedded between two normal/ferromagnetic leads ($L$ and $R$) and is side-attached to QD$_2$, which is coupled to  superconductor ($S$).}
\label{scheme}
\end{figure}

Charge transfer through multiterminal junctions is even more complex, because of several possible transport channels. For instance, three-terminal setup with the single quantum dot on interface of two normal and another superconducting electrode enables the single electron transfer (ET) as well as the direct (DAR) and/or crossed (CAR) Andreev reflections between the normal electrodes \cite{Michalek-2013, Weymann-2014, Wojcik-2014, Siqueira-2015, Michalek-2016, Gramich-2017, Pawlicki-2018, Hashimoto-2024}. Under such circumstances, both the local and nonlocal transport coefficients can be measured \cite{Mazza-2014}. It has been shown that when the crossed Andreev reflections prevails over the ballistic ET the nonlocal conductance acquires negative values  \cite{Pillet-2010,Andersen-2011,Gramich-2017,Higginbotham-2015,He-2020}. Furthermore, temperature difference imposed in such setup between the normal electrodes allows for separating the charge from the heat currents  \cite{Sanchez-2011,Thierschmann-2015,Szukiewicz-2015,Sothmann-2015}.
Interesting properties occur also in multiterminal geometries where QD is  between one normal and two superconducting electrodes, forming the Josephson-type junction \cite{Oguri-2012,Oguri-2013,Kirsanskas-2015,Domanski-2017}. The Kondo and Andreev effects can be there controlled by phase difference (via magnetic field) between the superconducting electrodes. 

In this paper we investigate signatures of the superconducting proximity effect appearing in the local/nonlocal transport properties of three-terminal junction, considering two quantum dots in T-shape geometry between two normal/ferromagnetic electrodes and superconducting lead \cite{Siqueira-2012,Calle-2013, Calle-2017,Wojcik-2018,Calle-2020,Gonzalez-2023} (see Fig.\ \ref{scheme}). Bound states of the double dot molecules have been so far probed experimentally in two-terminal junctions,  using the scanning tunneling technique \cite{Franke-2024,Zonda-2025} as well as the Josephson \cite{Paaske-2023} and the Andreev spectroscopy \cite{Frolov-2022}. Here we consider three-terminal configuration,  where the superconducting proximity effect is indirectly transmitted to one of the dots (QD$_1$) via the other one (QD$_2$), therefore the interdot coupling plays decisive role in affecting the local/nonlocal transport properties. 

From numerical calculations we find that in the weak interdot coupling regime the interferometric (Fano-type) features show up, whereas for the tightly coupled dots the molecular bound states give rise to the negative conductance and divergence of the Seebeck coefficient. These phenomena are caused solely by the Andreev-type channels. We investigate these effects, focusing on the linear response in the deep subgap region.
  
The paper is organized as follows. We start by introducing the microscopic model (Sec.\ \ref{sec:Hamiltonian}) and define the transport coefficients  (Secs.\ \ref{sec:Current} and  \ref{sec:currentLinear}). Next, we present our numerical results obtained for the normal (Sec. \ref{sec:normal}) and for the ferromagnetic electrodes (Sec.\ \ref{sec:polarLeads}), examining the local and nonlocal properties of charge conductance and thermopower. In  Sec.\ \ref{sec:summary} we summarize the paper. Role of the Coulomb interaction is briefly disscused in Appendix \ref{sec:correlations} and  influence of temperature on the transport coefficients of polarized system is presented in Appendix \ref{sec:coefficients}.

\section{Model}
\label{sec:model}

We consider the three-terminal junction with two quantum dots in T-shape geometry (Fig. \ref{scheme}), assuming the the central quantum dot (QD$_1$) to be weakly hybridized with the metallic (or ferromagnetic) electrodes, and the second quantum dot (QD$_2$)  strongly coupled to s-wave superconducting (S) lead. Such asymmetry of the couplings guarantees that proximity effect induces narrow in-gap states at QD$_2$ which (through the interdot coupling) affect the spectrum of QD$_1$, influencing the local and nonlocal transport properties of the junction.

\subsection{Hamiltonian}
\label{sec:Hamiltonian}

Our hybrid structure can be described by the following Hamiltonian
\begin{eqnarray}
H=H_{DQD} + H_{N-QD1} +H_{S-QD2} .
\label{HAM}
\end{eqnarray}
The double quantum dot term is given by
\bea
H_{DQD}=\sum_{i\sigma}\varepsilon_{i\sigma} d^{\dagger}_{i\sigma} d_{i\sigma}
+\sum_{\sigma}t_{12}(d^{\dagger}_{1\sigma} d_{2\sigma}+h.c.) ,  
\label{HDQD}
\eea
where operators $d_{i\sigma}^\dagger(d_{i\sigma})$ create (annihilate) electrons on the $i$-th quantum dot with spin $\sigma$ and energy $\varepsilon_{i\sigma}$. For numerical computations, we assume the energy levels $\varepsilon_{i\sigma}=\varepsilon_{i}$ and the interdot hopping $t_{12}$ to be spin-independent. This assumption is valid in absence of external magnetic field, otherwise  the Zeeman splitting shoud be taken into account \cite{Li-2008,Wang-2012,Yao-2018}.

The term describing the normal/ferromagnetic leads and their hybridization with  QD$_1$ can be expressed as
\bea
\label{HNQD}
H_{N-QD1} &=& \sum_{\bk,\sigma,\alpha} (\varepsilon_{\bk\alpha\sigma}-\mu_{\alpha})c^{\dagger}_{\bk\alpha\sigma}c_{\bk\alpha\sigma}\\
&+&\sum_{\bk,\sigma}(V_{\bk\alpha\sigma}d^{\dagger}_{1\sigma} c_{\bk\alpha\sigma}+h.c.), \nonumber
\eea
where $c^{\dagger}_{\bk\alpha\sigma}(c_{\bk\alpha\sigma})$ is the creation (annihilation) operator of spin $\sigma$ electron with momentum $\bk$ in $\alpha={L,R}$ lead, $\varepsilon_{\bk \alpha \sigma}$ is the kinetic energy, $\mu_{\alpha}$ denotes the chemical potential, and $V_{k\alpha}$ is the hopping between  external leads and QD$_1$. In the wide bandwidth limit, we can introduce the energy-independent tunnel couplings
$\Gamma_{\alpha\sigma}=2\pi\sum_{k}\left|V_{k \alpha\sigma}\right|^2\delta(\omega-\varepsilon_{\bk \alpha\sigma}+\mu_{\alpha\sigma})$. In Sec.\ \ref{sec:polarLeads} we consider their spin-polarized versions, $\Gamma_{\alpha\sigma}=\Gamma_{\alpha}(1+\sigma p_\alpha)$,  assuming that $p_L=p_R \equiv p_0$. 

The superconducting lead which is directly coupled to QD$_2$ will be treated within the BCS framework \cite{Yamada-2011}
\bea
\label{HSC}
H_{S-QD2} &=& \sum_{\bk,\sigma} (\varepsilon_{\bk S \sigma}-\mu_{S})c^{\dagger}_{\bk  S \sigma}c_{\bk S \sigma}\\
&+& \sum_{\bk} \Delta (c_{\bk  S \downarrow}c_{-\bk S \uparrow}+ c^{\dagger}_{-\bk  S \uparrow}c^{\dagger}_{\bk S \downarrow}) \nonumber\\
&+&\sum_{\bk,\sigma}(V_{\bk S\sigma}d^{\dagger}_{2\sigma} c_{\bk S\sigma}+h.c.), \nonumber
\eea
where again the operator $c^{\dagger}_{\bk S \sigma}(c_{\bk S \sigma})$ refer to creation (annihilation) of spin $\sigma$ electron with momentum $\bk$, the kinetic energy $\varepsilon_{\bk S \sigma}$ is measured with respect to the chemical potential $\mu_S$ and $\Delta$ denotes the isotropic pairing gap.
For convenience, we assume the superconducting lead to be grounded, $\mu_{S}=0$.

In absence of the interdot coupling ($t_{12}=0$) and in the superconducting atomic limit ($\Delta\rightarrow \infty$), the spectrum of QD$_2$ is characterized by a pair of the Adreev bound states at energies $E_{A\pm}=\pm\sqrt{\varepsilon_2^2+(\Gamma_S/2)^2}$, where  $\Gamma_{S}=2\pi\sum_{\bk}\left|V_{\bk S\sigma}\right|^2\delta(\omega-\varepsilon_{{\bk S}\sigma}+\mu_{S})$ is the coupling strength between QD$_2$ and superconducting lead.
These Andreev states hybridize with the energy level of QD$_1$ through the interdot coupling, $t_{12}$, leading to development of the molecular structure of the double quantum dot. For the uncorrelated setup we obtain the effective quasiparticle states at energies
\begin{eqnarray}
\varepsilon_{AD1}^{\pm}=\pm \frac{1}{\sqrt{2}} \sqrt{A - \sqrt{A^2-4B}}, 
\label{eq_6}\\
\varepsilon_{AD2}^{\pm}=\pm \frac{1}{\sqrt{2}} \sqrt{A + \sqrt{A^2-4B}} ,
\label{eq_7}
\end{eqnarray}
where 
\begin{eqnarray}
A&=&\varepsilon_1^2+E^2_{A+}+2t_{12}^2, \\
B&=&(\varepsilon_1\varepsilon_2-t_{12}^2)^2+\left(\varepsilon_1\Gamma_{S}/2\right)^2 .
\end{eqnarray}
%
Spectrum of the central quantum dot 
is presented in Secs.\  \ref{sec:normal}  and  \ref{sec:polarLeads}. For $t_{12}\rightarrow 0$ the quasiparticle energies simplify to $\varepsilon_{AD1}^{\pm}\rightarrow \varepsilon_1$ and $\varepsilon_{AD2}^{\pm}\rightarrow E_{A\pm}$, respectively.

Transport properties of this three-terminal system induced by the voltage applied to the normal leads $\mu_{\alpha}=eV_\alpha$ and/or by temperature difference $T_L\neq T_R$  depend on the effective quasiparticle states of the quantum dots. In what follows we provide specific details, concerning this issue.

\subsection{Charge transport}
\label{sec:Current}

Charge current from the $L$-th lead can be expressed by
\bea
\label{ILdef}
J_{L\sigma} &=& -e\frac{d}{dt}\left\langle \sum_{\bk}c^{\dagger}_{\bk L \sigma}c_{\bk L \sigma}\right\rangle\\
&=&\frac{ie}{\hbar}\left\langle \sum_{\bk}\left[c^{\dagger}_{\bk L \sigma}c_{\bk L \sigma},H\right]_-\right\rangle. \nonumber
\eea
From a straightforward analysis we obtain
\bea
\label{ILdef2}
J_{L\sigma} =\frac{2e}{\hbar} \sum_{\bk} \Re\left[V_{\bk L \sigma}{\cal{G}}_{1\sigma,\bk L \sigma}^<(t,t)\right]
\eea
where ${\cal{G}}_{1\sigma,\bk L \sigma}^<(t,t')=i\left\langle c^{\dagger}_{\bk L \sigma}(t')d_{1\sigma}(t) \right\rangle$ is the lesser Green's function. Introducing the Fourier transform and following the procedure formulated by Haug and Jauho \cite{Haug-2008} we can express $J_{L\sigma}$ as
\bea
\label{ILdef3}
J_{L\sigma} &=&-\frac{2e}{h} \frac{\Gamma_{L\sigma}}{2}\int d\omega\Im\left[2f_{L\sigma}\left\langle \left\langle d_{1\sigma}\left|d^\dagger_{1\sigma}\right.\right\rangle\right\rangle^r_\omega\right.\\
&+&\left.\left\langle \left\langle d_{1\sigma}\left|d^\dagger_{1\sigma}\right.\right\rangle\right\rangle^<_\omega\right],\nonumber
\eea
where $f_{\alpha\sigma}=\left\{\text{exp}\left[(\omega-\mu_{\alpha\sigma})/k_B T_{\alpha}\right]+1\right\}^{-1}$ is the Fermi-Dirac distribution function.

The retarded Green's function $\left\langle \left\langle d_{1\sigma}\left|d^\dagger_{1\sigma}\right.\right\rangle\right\rangle^r_\omega$ can be determined from the equation of motion 
\begin{eqnarray}
\omega\left\langle \left\langle \Psi_i\left|\Psi_j\right.\right\rangle\right\rangle^r_\omega= \left\langle \left[\Psi_i,\Psi_j\right]_+ \right\rangle + \left\langle \left\langle \left[\Psi_i,H\right]_-\left|\Psi_j\right.\right\rangle\right\rangle^r_\omega ,
\label{EOM}
\end{eqnarray}
for the matrix Green's function $\hat {\cal{G}}^r(\omega )=\left\langle \left\langle\hat{\Psi} \left|\hat{\Psi}^\dag\right.\right\rangle\right\rangle _\omega ^r$ which is defined in the Nambu spinor representation $\hat{\Psi}^\dag  = (d_{1\uparrow}^\dag,d_{1\downarrow},d_{2\uparrow}^\dag,d_{2\downarrow})$. 
On the other hand, the lesser Green's function 
obeys the Keldysh equation \cite{Haug-2008}
\begin{eqnarray}
{\hat{\cal{G}}}^<(\omega)={\hat{\cal{G}}}^r(\omega)\hat{\Sigma}^<(\omega){\hat{\cal{G}}}^a(\omega) ,
\label{Keldysh}
\end{eqnarray}
where $\hat{\Sigma}^<(\omega)$ denotes the lesser self-energy matrix.

The lesser self-energy matrix, appearing in Eqn.\ (\ref{Keldysh}), can be expressed as
\begin{widetext}
\begin{eqnarray} 
\hat{\Sigma}^<(\omega) =
-i\left( \begin{array}{cccc}  
\Gamma_{L\uparrow}f_{L\uparrow}+\Gamma_{R\uparrow}f_{R\uparrow}&0&0&0\\
0&\Gamma_{L\downarrow}\tilde{f}_{L\downarrow}+\Gamma_{R\downarrow}\tilde{f}_{R\downarrow}&0& 0\\
0&0&\Gamma_{S}\beta(\omega)f_{S} &\Gamma_{S}\beta(\omega)\frac{\Delta}{\omega}f_{S}\\
0& 0 & \Gamma_{S}\beta(\omega)\frac{\Delta^*}{\omega}f_{S}&\Gamma_{S}\beta(\omega)f_{S}
\end{array}\right),
\label{SigLesser}
\end{eqnarray} 
%
where $\tilde{f}_{\alpha\sigma}=\left\{\text{exp}\left[(\omega+\mu_{\alpha\sigma})/k_B T_{\alpha}\right]+1\right\}^{-1}$ is the distribution function for holes and $\beta(\omega)=\frac{\left|\omega\right|\Theta(\left|\omega\right|-\Delta)}{\sqrt{\omega^2-\Delta^2}}-\frac{i\omega\Theta(\Delta-\left|\omega\right|)}{\sqrt{\Delta^2-\omega^2}}$. 
For the uncorrelated setup (neglecting the Coulomb repulsion on both quantum dots) we obtain  \cite{Yamada-2011}
%
\begin{eqnarray} 
{\hat{\cal{G}}}^{r(a)}(\omega) =
\left( \begin{array}{cccc}  
\omega-\epsilon_{1\uparrow}\pm i\frac{\Gamma_{N \uparrow}}{2} &0&t_{12}&0\\
0&\omega+\epsilon_{1\downarrow}\pm i\frac{\Gamma_{N \downarrow}}{2}&0& -t_{12}\\
t_{12}&0&\omega-\epsilon_{2\uparrow}\pm i\frac{\Gamma_{S}}{2}\beta(\omega) &\pm i\frac{\Gamma_{S}}{2}\beta(\omega)\frac{\Delta}{\omega}\\
0& -t_{12} &\pm i\frac{\Gamma_{S}}{2}\beta(\omega)\frac{\Delta^*}{\omega}& \omega+\epsilon_{2\downarrow}\pm i\frac{\Gamma_{S}}{2}\beta(\omega)
\end{array}\right)^{-1} .
\label{Gr44}
\end{eqnarray} 
\end{widetext}

\begin{figure*}
\begin{center}
\includegraphics[width=1\linewidth]{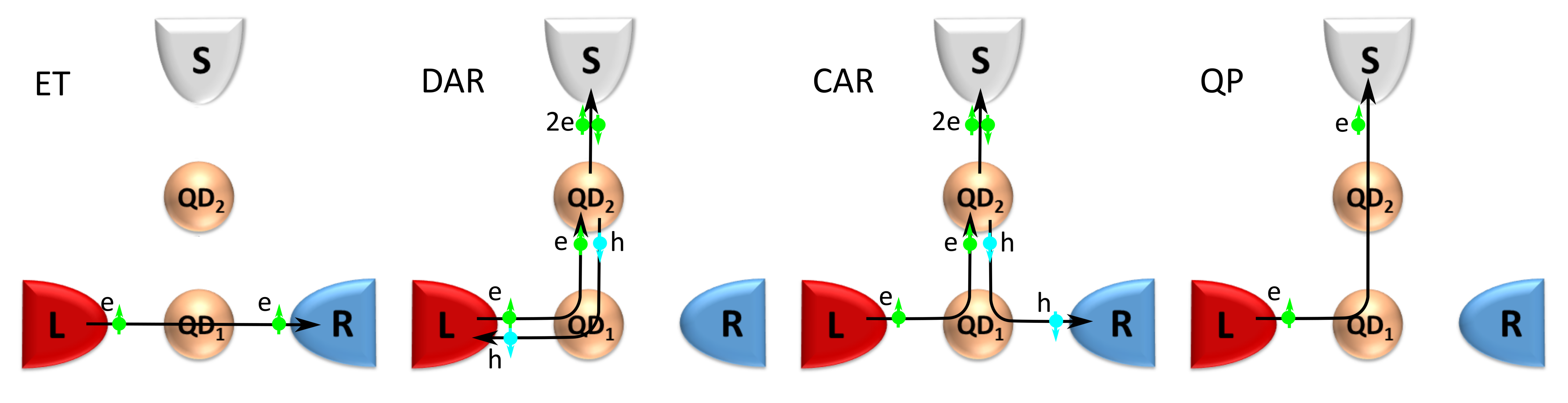}
\end{center}
\caption{Illustration of the charge transport processes in the three-terminal setup contributed by: ballistic electron transfer (ET), direct Andreev reflection (DAR), crossed Andreev reflection (CAR) and  quasiparticles tunneling (QP), respectively.}
\label{transport_channels}
\end{figure*}

Using this formalism we can represent the charge current (\ref{ILdef3}) by contributions from the ballistic electron transfer (ET), the direct (DAR) and crossed (CAR) Andreev reflections, and the quasiparticle flow (QP)
\bea
\label{Jstot}
J_{L\sigma}= J^{ET}_{L\sigma}+J^{DAR}_{L\sigma} +J^{CAR}_{L\sigma}+J^{QP}_{L\sigma} .
\eea
These transport channels are graphically displayed in Fig.\ \ref{transport_channels}.
The ballistic transfer of electrons from $L$ to $R$ lead through QD$_1$ can be expressed as
\bea
\label{JsET}
J^{ET}_{L\sigma}= \frac{e}{h}\int{d\omega T^{ET}_{\sigma}(\omega)\left[f_{L\sigma}(\omega)-f_{R\sigma}(\omega)\right]},
 \eea
where the tunneling transmittance is given by
\bea
\label{TET}
T^{ET}_{\sigma}(\omega)&=&\Gamma_{L\sigma}\Gamma_{R\sigma}\left|\left\langle \left\langle d_{1\sigma}\left|d_{1\sigma}^\dag\right.\right\rangle\right\rangle^r_\omega\right|^2.
\eea
The direct Andreev reflection  (DAR) describes a conversion of electron from the $L$-th lead into the local pair at QD$_2$ which is next injected to the superconductor, while a hole of opposite spin is scattered back to the same electrode. Such current is given by
\bea
\label{JsDAR}
J^{DAR}_{L\sigma}=\frac{e}{\hbar}\int{d\omega T^{DAR}_{\sigma}(\omega)\left[f_{L\sigma}(\omega)-\tilde{f}_{L-\sigma}(\omega)\right]},
\eea
with the transmittance
\bea
\label{TDAR}
T^{DAR}_{\sigma}(\omega)=\Gamma_{L\sigma}\Gamma_{L-\sigma}\left|\left\langle \left\langle d_{1\sigma}\left|d_{1-\sigma}\right.\right\rangle\right\rangle^r_\omega\right|^2.
\eea
The crossed Andreev reflection process is similar to DAR, except that it injects a hole to the $R$-th lead
\bea
\label{JsCAR}
J^{CAR}_{L\sigma}&=&\frac{e}{\hbar}\int{d\omega T^{CAR}_{\sigma}(\omega)\left[f_{L\sigma}(\omega)-\tilde{f}_{R-\sigma}(\omega)\right]},
\eea
where the corresponding transmittance
\bea
\label{TCAR}
T^{CAR}_{\sigma}(\omega)&=&\Gamma_{L\sigma}\Gamma_{R-\sigma}\left|\left\langle \left\langle d_{1\sigma}\left|d_{1-\sigma}\right.\right\rangle\right\rangle^r_\omega\right|^2.
\eea
The last term in Eqn.\ (\ref{Jstot}) describes the single electron transfer from the $L$-th lead to a continuum outside the pairing gap of superconductor. This quasiparticle current is given as
\bea
\label{JsQP}
J^{QP}_{L\sigma}&=& \frac{e}{\hbar}\int{d\omega T^{QP}_{\sigma}(\omega)\left[f_{L\sigma}(\omega)-f_{S\sigma}(\omega)\right]}
\eea
with the transmittance \cite{Yamada-2010}
\bea
\label{TQP}
T^{QP}_{\sigma}(\omega)&=&\Re \left[\beta(\omega)\right]\Gamma_{L\sigma}\Gamma_{S\sigma}\left(\left|\left\langle \left\langle d_{1\sigma}\left|d_{2\sigma}^\dagger\right.\right\rangle\right\rangle^r_\omega\right|^2\right.\\
&+&\left|\left\langle \left\langle d_{1\sigma}\left|d_{2-\sigma}\right.\right\rangle\right\rangle^r_\omega\right|^2\nonumber\\
&+&\left.2\frac{\Delta}{\omega}\Re{\left[\left\langle \left\langle d_{1\sigma}\left|d_{2\sigma}^\dagger\right.\right\rangle\right\rangle^r_\omega \left\langle \left\langle d_{2-\sigma}^\dagger\left|d_{1\sigma}^\dagger\right.\right\rangle\right\rangle^a_\omega\right]}\right).\nonumber
\label{T_QP}
\eea
Let us notice that Eqn.\ (\ref{T_QP}) depends on $\Re \left[\beta(\omega)\right]$ whose nonvanishing value $T^{QP}_{\sigma}(\omega)$ occurs solely for $|\omega|>\Delta$. Such current $J^{QP}$ is hence meaningful only for suffinciently large voltage, $\left|eV_L\right|>\Delta$, and/or at sufficiently large temperatures, $k_{B}T \geq \Delta$.

In analogous way we can express the charge current from $R$-th electrode,  $J_R$, by swapping the indices $L\leftrightarrow R$.

\subsection{Linear response limit}
\label{sec:currentLinear}

In the present work we shall focus on the low temperature limit $T \ll T_c$ (where $T_c$ denotes the superconducting critical temperature) and consider infinitesimally small perturbations, $\mu_\alpha=e\delta V_\alpha$ and $T_{L,R}=T_{S}+\delta T_{L,R}$. The charge current from $L$-th lead simplifies then to
 \bea
\label{JsLtotdel}
J_{L\sigma}&=& D_{\sigma\mu}^{ET}(e\delta V_L-e\delta V_R)+
2D_{\sigma\mu}^{DAR}e\delta V_L \\
&+&D_{\sigma\mu}^{CAR}(e\delta V_L+e\delta V_R)+D_{\sigma\mu}^{QP}e\delta V_L\nn \\
&+&D_{\sigma T}^{QP}\delta T_L+(D_{\sigma T}^{ET}+D_{\sigma T}^{CAR})(\delta T_L-\delta T_R) \nn ,
\eea
where the coefficients
\bea
\label{DLRmu}
D_{\sigma\mu}^{\kappa}&=&\frac{e}{\hbar}\int {d\omega T^{\kappa}_{\sigma}(\omega)\left[-\displaystyle \frac{\partial {f}}{\partial \omega}\right]},
\\
\label{DLRT}
D_{\sigma T}^{\kappa}&=&\frac{e}{\hbar T}\int {d\omega \omega T^{\kappa}_{\sigma}(\omega)\left[-\displaystyle \frac{\partial f}{\partial \omega}\right]}
\eea
refer to $\kappa=\left\{\text{ET},\text{DAR},\text{CAR},\text{QP}\right\}$ transport channels.

Let us now define the local $G_{\alpha\alpha\sigma}= \left.{\frac{d J_{\alpha\sigma}}{d V_\alpha}}\right|_{\delta T=0}$ and nonlocal electrical conductance $G_{\alpha\beta\sigma}= -\left.{\frac{d J_{\alpha\sigma}}{d V_\beta}}\right|_{\delta T=0}$, where $\alpha\neq\beta$ \cite{Mazza-2014,Michalek-2013,Michalek-2016}.
Eqn.\ (\ref{Jstot}) implies that the local  conductance is contributed from all transport channels 
\bea
\label{GLL}
G_{LL\sigma}&=&G_{\sigma}^{ET}+2G_{\sigma}^{DAR}+G_{\sigma}^{CAR}+G_{\sigma}^{QP} ,
\eea
whereas the nonlocal conductance is given by a difference of the electron transfer and the crossed Andreev reflection 
\bea
G_{RL\sigma}&=&G_{\sigma}^{ET} - G_{\sigma}^{CAR},
\label{eq_18}
\eea
where $G_{\sigma}^\kappa=eD_{\sigma\mu}^{\kappa}$.
We clearly notice that the nonlocal conductance  (\ref{eq_18}) takes either positive or negative values, depending on the dominant transport channel. In particular, it would be negative when the superconducting proximity effect plays major role promoting the Andreev scattering over the single electron ballistic transfer.

At zero temperature the transport coefficients $D_{\sigma\mu}^{\kappa}$ of our uncorrelated setup simplify to
\begin{equation}
D_{\sigma\mu}^{ET}=\frac{e}{\hbar}\frac{\Gamma_{L\sigma}\Gamma_{R\sigma}\left[\left(\en_2 t_{12}^2-\en_1 E^{2}_{A+}\right)^2+\frac{\Gamma_N^2}{4}E^{4}_{A+}\right]}{\left[\frac{\Gamma_N^2}{4}E^{2}_{A+}+B\right]^2},
\label{DETT0}
\end{equation}
\bea
D_{\sigma\mu}^{DAR}&=&\frac{e}{\hbar}\frac{\Gamma_{L\sigma}\Gamma_{L-\sigma}\frac{\Gamma_S^2}{4}t_{12}^4}{\left[\frac{\Gamma_N^2}{4}E^{2}_{A+}+B\right]^2} ,
\label{DDART0}
\\
D_{\sigma\mu}^{CAR}&=&\frac{e}{\hbar}\frac{\Gamma_{L\sigma}\Gamma_{R-\sigma}\frac{\Gamma_S^2}{4}t_{12}^4}{\left[\frac{\Gamma_N^2}{4}E^{2}_{A+}+B\right]^2}
\label{DCART0}
\eea
and
\bea
D_{\sigma\mu}^{QP}=0.
\label{DQPT0}
\eea
Under such circumstances, the coefficients $D_{\sigma\mu}^{\kappa}$ do not depend on the pairing gap of superconductor.

From Eqs.\ (\ref{DETT0}-\ref{DQPT0}) we can derive explicit expressions for the local and nonlocal electrical conductance
\begin{widetext}
\begin{eqnarray}
G_{LL\sigma}&=&\frac{e^2}{\hbar}\frac{\Gamma_{L\sigma}\Gamma_{R\sigma}\left[\left(\en_2 t_{12}^2-\en_1 E^{2}_{A+}\right)^2+\frac{\Gamma_N^2}{4}E^{4}_{A+}\right]+2\Gamma_{L\sigma}\Gamma_{L-\sigma}\frac{\Gamma_S^2}{4}t_{12}^4+\Gamma_{L\sigma}\Gamma_{R-\sigma}\frac{\Gamma_S^2}{4}t_{12}^4}{\left[\frac{\Gamma_N^2}{4}E^{2}_{A+}+B\right]^2}
\label{GLLT0}
\\
G_{RL\sigma}&=&\frac{e^2}{\hbar}\frac{\Gamma_{L\sigma}\Gamma_{R\sigma}\left[\left(\en_2 t_{12}^2-\en_1 E^{2}_{A+}\right)^2+\frac{\Gamma_N^2}{4}E^{4}_{A+}\right]-\Gamma_{L\sigma}\Gamma_{R-\sigma}\frac{\Gamma_S^2}{4}t_{12}^4}{\left[\frac{\Gamma_N^2}{4}E^{2}_{A+}+B\right]^2}.
\label{GRLT0}
\end{eqnarray}
\end{widetext}

For the setup with ferromagnetic electrodes the conductance would depend on spin, $G_{LL\up} \neq G_{LL\down}$.  Under such conditions the local and nonlocal conductance is
\bea
\label{GLLGRLTOT}
G_{LL}=\sum_{\sigma}G_{LL\sigma}  \hspace{0.3cm}\text{and} 
\hspace{0.3cm} G_{RL}=\sum_{\sigma}G_{RL\sigma}.
\eea
We can also introduce their spin-polarized versions
\bea
\label{PGLLGRL}
P_{G_{LL}}=\frac{G_{LL\up} - G_{LL\down}}{G_{LL}}, 
\hspace{0.2cm}
P_{G_{RL}}=\frac{G_{RL\up} - G_{RL\down}}{G_{RL}}.
\eea
 
In similar way, for the temperature gradient $\Delta T_\alpha$ we define the local $S_{\alpha\alpha\sigma}= -\left.{\frac{d V_\alpha}{d T_\alpha}}\right|_{J_{\gamma\sigma}=0}$ and nonlocal $ S_{\alpha\beta\sigma}=- \left.{\frac{d V_\alpha}{d T_\beta}}\right|_{J_{\gamma\sigma}=0} $ Seebeck coefficients where voltage $V_{\alpha}$ compensates the current induced by temperature gradient, $J_{\gamma\sigma}=0$.  For infinitesimally small temperature difference the thermopower can be expressed by
\bea
\label{SLLsigma}
S_{LL\sigma}&=&\frac{D_{\sigma T}^{ET}+D_{\sigma T}^{CAR}+D_{\sigma T}^{QP}}{G_{LL\sigma}} ,
\\
\label{SRLsigma}
S_{RL\sigma}&=&-\frac{D_{\sigma T}^{ET}+D_{\sigma T}^{CAR}+D_{\sigma T}^{QP}}{G_{RL\sigma}}.
\eea

In the case of ferromagnetic electrodes $S_{LL\up} \neq S_{LL\down}$,  therefore we can consider the averaged Seebeck coeeficients
\bea
\label{eq:S}
{S_{LL}}=\frac{S_{LL\up} + S_{LL\down}}{2} \text{   and   } {S_{RL}}=\frac{S_{RL\up} + S_{RL\down}}{2} 
\eea
and their spin-resolved counterparts \cite{Swirkowicz-2009,Weymann-2013}
\bea
\label{eq:SS}
{S^S_{LL}}=\frac{S_{LL\up} - S_{LL\down}}{2}, \hspace{0.1cm}
S^S_{RL}=\frac{S_{RL\up} - S_{RL\down}}{2}.
\eea

In the next sections we present the characteristic properties obtained numerically for of the local and nonlocal transport coefficients of three-terminal setup with the normal and ferromagnetic leads, respectively.

\section{Setup with normal electrodes}
\label{sec:normal}

Let us start by considering the spin-resolved spectrum of the central quantum dot (QD$_{1}$) and next discuss its influence on  the transport properties of our system. For computations, we treat $\Gamma_N=\Gamma_{L}+\Gamma_{R}\equiv 1$ as a convenient energy unit and assume the strong coupling to superconductor $\Gamma_S=2\Gamma_{N}$ to obtain well pronounced signatures of the quasiparticle states in the spectra of both quantum dots. 

Eqs.\ (\ref{GLLT0}) and (\ref{GRLT0}) show that at zero temperature the local and nonlocal conductance do not depend on the pairing gap of superconducting lead. For this reason, to simplify our study and reduce a number of the model parameters, we restrict to the superconducting atomic limit approach ($\Delta\rightarrow \infty$).
This approach is valid when the energies $\varepsilon_{i}$ are deep inside the pairing gap of superconductor, $\Delta$, and temperature is safely smaller than $T_c$. Otherwise, the continuum electronic states from outside the pairing would play important role.

\subsection{Quasiparticle spectrum}

\begin{figure}[tb] 
\centering
\includegraphics[width=1\linewidth]{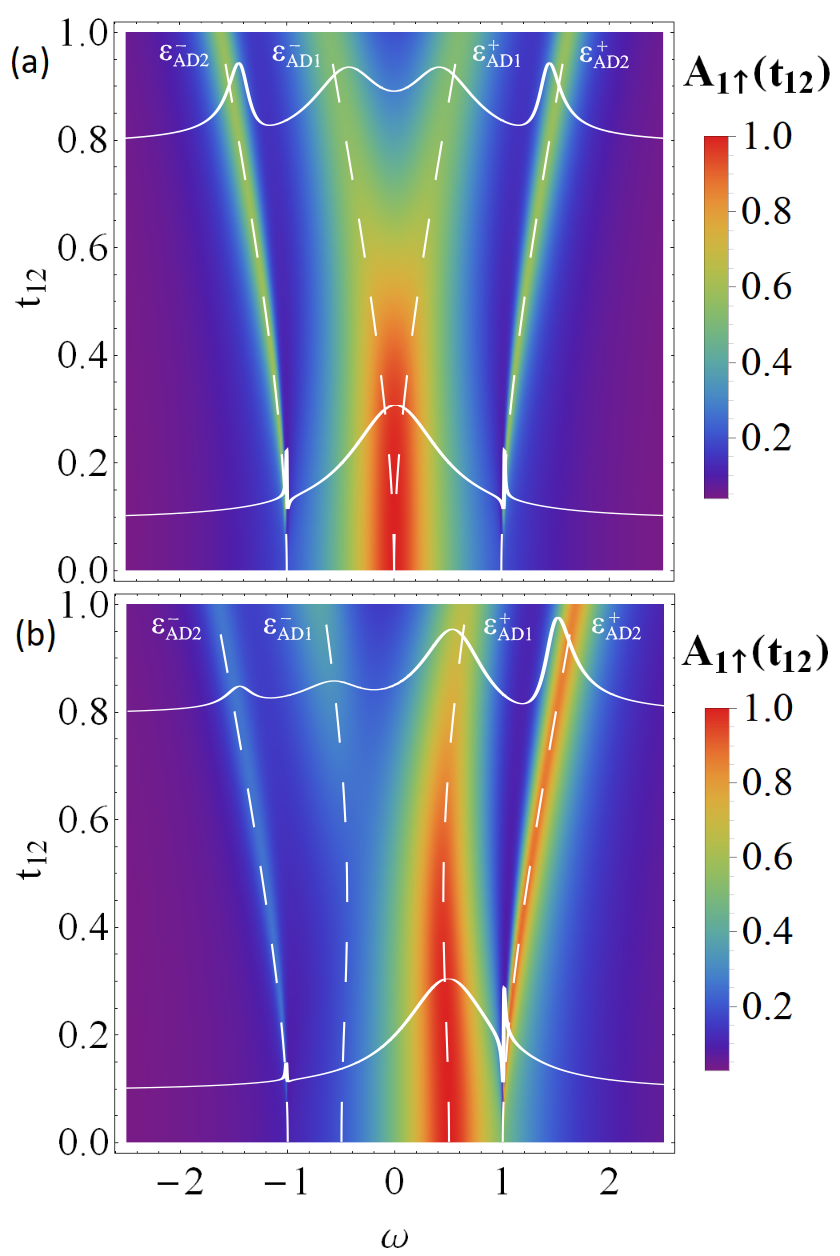}
\caption{
The spectral function $A_{1\uparrow}(\omega)$ of QD$_1$ as a function of the interdot coupling $t_{12}$ obtained for $\varepsilon_{1}=0$ (a) and $\varepsilon_{1}=0.5$ (b) and using the model parameters $\varepsilon_{2}=0$, $\Gamma_{S}=2$, $\Gamma_{L}=\Gamma_{R}=0.5$, $k_BT=0$ and $p_{0}=0$. White solid-lines display the profile of the spectral function in the weak ($t_{12}=0.1$) and strong  coupling ($t_{12}=0.8$) limits, respectively. The quasiparticle energies $\varepsilon_{AD1}^{\pm}$ and $\varepsilon_{AD2}^{\pm}$, given by Eqs.\ (\ref{eq_6},\ref{eq_7}), are marked by white-dashed lines.}
\label{figA1kt12}
\end{figure}

We first analyze the  quasiparticles of the quantum dots, depending on the interdot coupling and their energy levels. Let us consider the spectral function
\begin{equation}
A_{1\s}(\omega)=-\frac{\Gamma_N}{2}\mbox{\rm Im}\langle \langle  d_{1\s} ;d_{1\s}^\dag \rangle \rangle^r_{\omega}
\label{eq_33}
\end{equation}
 of the central quantum dot (normalized here by factor $\pi\Gamma_{N}/2$ to obtain dimensionless values). This quantum dot affects all four transport processes in our setup (displayed in Fig.\ \ref{transport_channels}). For the nonmagnetic L and R leads, this function (\ref{eq_33}) is spin-independent.

\begin{figure}[b!]
\centering
\includegraphics[width=0.8\linewidth]{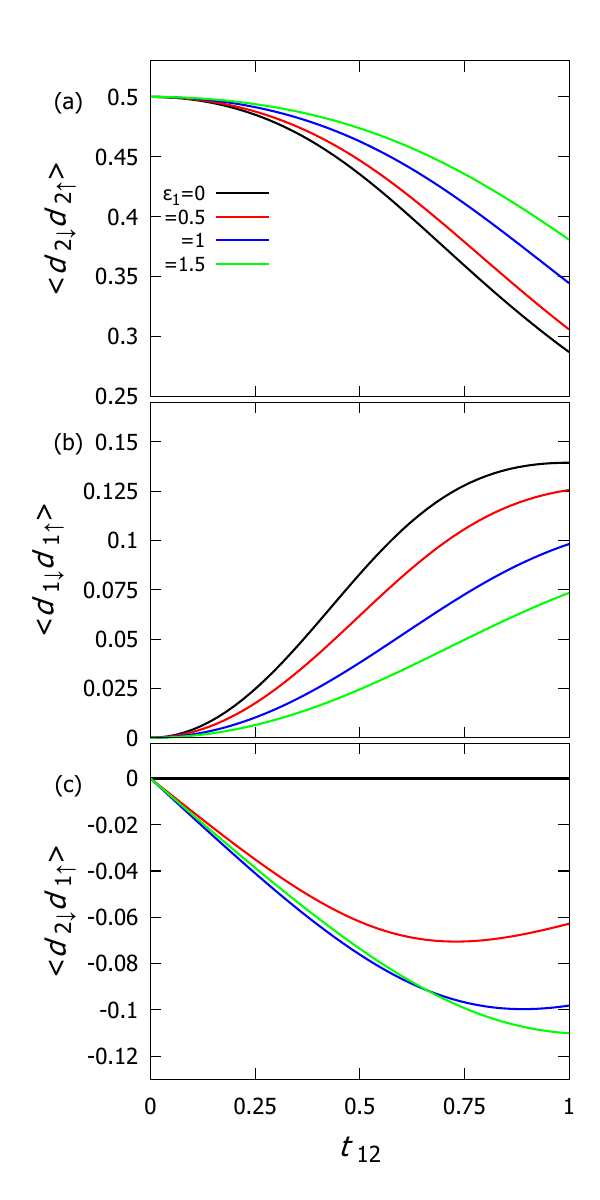}
\caption{
The local (panels a and b), and nonlocal (panel c) electron pairing plotted vs the interdot coupling $t_{12}$ for several energy levels $\varepsilon_{1}$, as indicated. Computations are done for the set of model parameters:  $\varepsilon_{2}=0$, $\Gamma_{S}=2$, $\Gamma_{L}=\Gamma_{R}=0.5$, $T=0$ and $p_{0}=0$.
} 
\label{figpair}
\end{figure}

In Fig.\ \ref{figA1kt12} we present the spectrum of $\uparrow$-spin electrons of QD$_1$ obtained for the energy level $\varepsilon_1=0$ (panel a) and $\varepsilon_1=0.5$ (panel b), respectively. 
 For isolated quantum dots, $t_{12}=0$, the spectral function of QD$_1$ takes the well-known Lorentzian shape centered near $\omega=\varepsilon_1$. For finite interdot coupling, $t_{12}\neq 0$,  we can observe two different effects, depending on the ratio $t_{12}/\Gamma_N$ \cite{Gonzalez-2023}. In the weak coupling limit, $t_{12}/\Gamma_N<0.5$, the interferometric (Fano-type) structures emerge at $\omega= E_{A\pm}$. In the strong coupling limit, $t_{12}/\Gamma_N\geq 0.5$, we observe development of the molecular quasiparticle spectrum characterized by four peaks (two pairs symmetrically placed around the Fermi level). This molecular structure is induced by leakage of the electron pairs from the superconducting lead onto both quantum dots. To support this claim we show in Fig.\ \ref{figpair} variation of the on-dot and inter-dot pairings
\begin{equation}
\left\langle d_{i\down}d_{j\up}\right\rangle=\frac{-1}{\pi}\int^{\infty}_{-\infty} \!\! \Imag \left\langle\left\langle d_{j\up};d_{i\down}\right\rangle\right\rangle_{\omega}^r 
\frac{1}{1+\e^{\omega/k_{B}T}} d\omega .
\label{ondot_pairing}
\end{equation}

Figure \ref{figpair} shows variation of the local and nonlocal pairings with respect to $t_{12}$ for several values of $\varepsilon_{1}$. Upon increasing $t_{12}$ the local pairing induced in QD$_1$ is gradually amplified at expense of a partial weakening of the electron pairing in QD$_2$. For the fully symmetric case,  $\varepsilon_{1}=0=\varepsilon_{2}$, the nonlocal singlet pairing $\left\langle d_{2\down}d_{1\up}\right\rangle$ is completely absent. It emerges solely in the asymmetric case, $\varepsilon_{1}\neq\varepsilon_{2}$, yet being an order of magnitude smaller in comparison to the local pairings. Of particular importance for the charge transport properties (discussed in next subsection) would be $\left\langle d_{1\down}d_{1\up}\right\rangle$ because it controls efficiency of the Andreev scattering processes. 

\begin{figure}[tb] 
\centering
\includegraphics[width=1\linewidth]{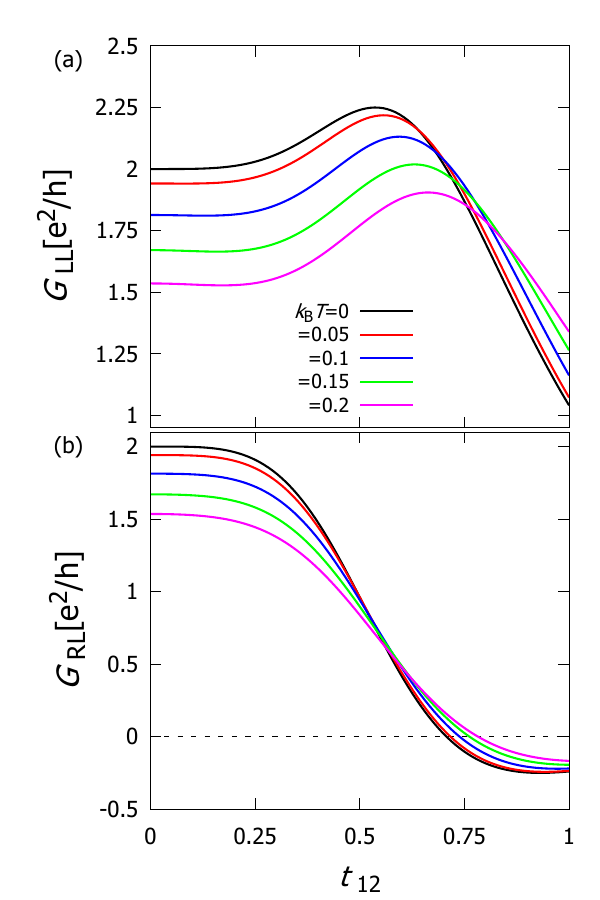}
\caption{
Variation of the local $G_{LL}$ (a) and nonlocal $G_{RL}$ (b) conductance  versus the interdot coupling $t_{12}$ for several temperatures, as indicated. Results are obtained for the model parameters $\varepsilon_{1}=\varepsilon_{2}=0$, $\Gamma_{S}=2$, $\Gamma_{L}=\Gamma_{R}=0.5$ and $p_{0}=0$.}
\label{figGLLGRL}
\end{figure}

\begin{figure}[tb] 
\centering
\includegraphics[width=1\linewidth]{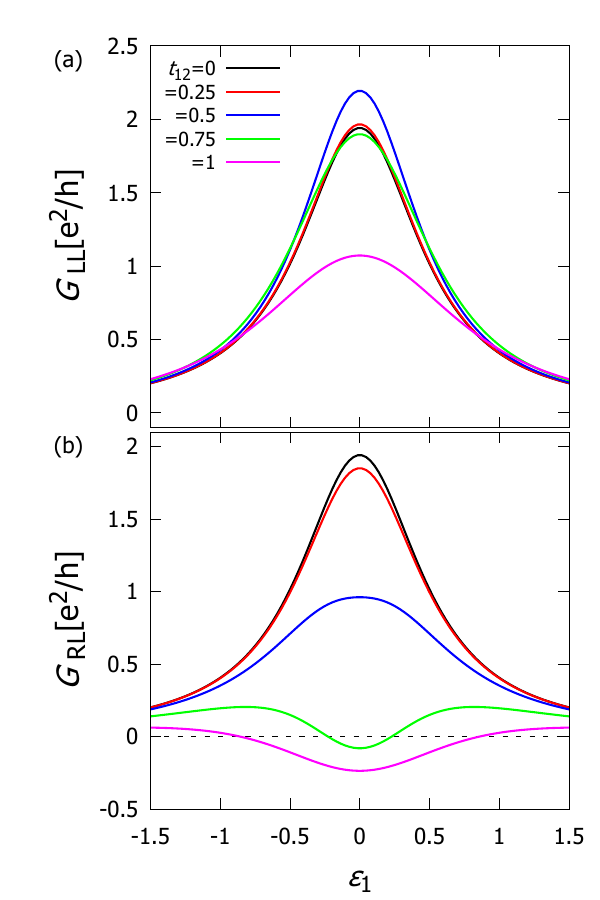}
\caption{
Local $G_{LL}$ (a) and nonlocal $G_{RL}$ (b) total zero-bias conductivity with respect to the QD$_1$ energy for different values of $t_{12}$. Other parameters are: $\varepsilon_{2}=0$, $\Gamma_{S}=2$, $\Gamma_{L}=\Gamma_{R}=0.5$, $p_{0}=0$} and $k_BT=0.05$.
\label{figGvEQD1kt12}
\end{figure}

\subsubsection{Zero-bias conductance}

Electron pairing induced in QD$_2$ by the proximity effect and subsequently transmitted to the central quantum dot, leads to characteristic features in the local and nonlocal transport properties. We discuss their signatures in the conductance and analyze their qualitative changes upon varying the interdot coupling, Fig.\  \ref{figGLLGRL}.

When the quantum dots are separated, $t_{12}=0$, the conductance is contributed only by the ballistic electron transport (\ref{DETT0}) which at zero temperature takes a unitary limit value $2e^{2}/h$.  By coupling these dots, $t_{12}\neq 0$, we observe slight reduction of $G^{ET}$ (due to suppression of the quasiparticle state at $\omega=0$) while  additional contribution comes from the Andreev  channel $G^{DAR(CAR)}$, enhancing the total conductance above its initial value $2e^2/h$. In the strong interdot coupling limit, $t_{12}>0.5$, the local conductance is substantially suppressed because the quasiparticle state, initially existing at $\omega=\varepsilon_1$, gradually evolves into  the molecular structure represented by four peaks (see Fig.\ \ref{figA1kt12}). 

In contrast to $G_{LL}$, the nonlocal conductance turns out to be very sensitive probe of the competition between the crossed Andreev reflections and the ballistic electron transfers. This property can be inferred from Eqn.\ (\ref{eq_18}) and can be observed for arbitrary interdot couplings $t_{12}$. 
Ultimately, for the tightly coupled quantum dots ($t_{12}\geq 0.75$) the sign-reversal of $G_{RL}$ occurs in analogy to the results previously obtained for three-terminal structure with the single quantum dot, both in the uncorrelated \cite{Michalek-2015} and strongly correlated \cite{Hashimoto-2024} cases.

\begin{figure}[tb] 
\centering
\includegraphics[width=1\linewidth]{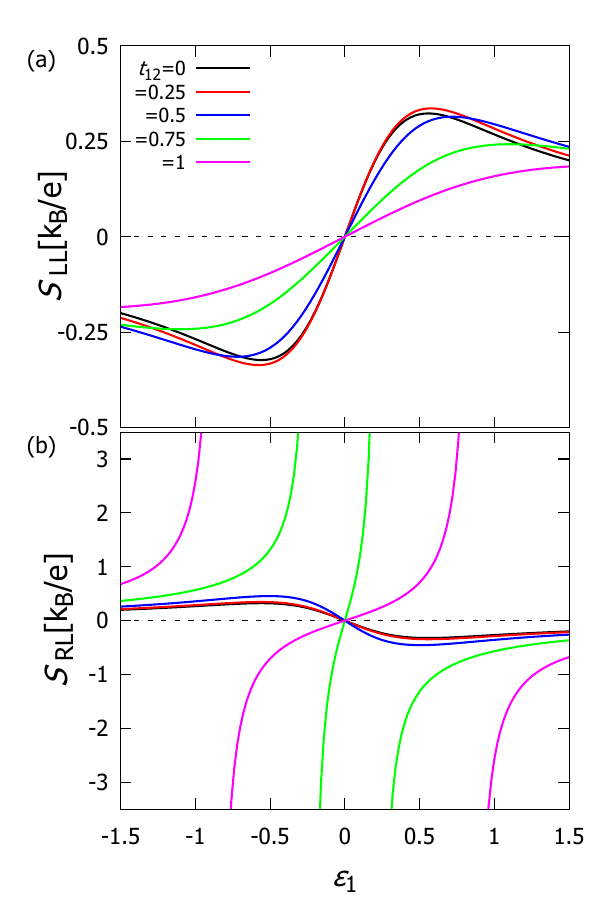}
\caption{
Local $S_{LL}$ (a) and nonlocal $S_{RL}$ (b) thermopower with respect to the QD$_1$ energy for different values of $t_{12}$,  using the model parameters: $\varepsilon_{2}=0$, $\Gamma_{S}=2$, $\Gamma_{L}=\Gamma_{R}=0.5$, $p_{0}=0$ and $k_BT=0.05$.} 
\label{figSvEQD1kt12}
\end{figure}

The energy levels have also some influence on efficiency of the superconducting proximity effect. Optimal conditions for the on-dot pairing induced in QD$_1$ occur in the symmetric case, $\varepsilon_{1,2}=0$ (black line in Fig.\ \ref{figpair}b). Under such circumstances the local conductance $G_{LL}$  reaches its maximum  at $t_{12}\approx 0.5$ (see Fig.\ \ref{figGvEQD1kt12}a). 
Far away the symmetric case the on-dot pairing (\ref{ondot_pairing}) becomes residual, therefore  the local conductance stays practically does not vary against $\varepsilon_1$. This behavior is evident from Fig.\ \ref{figGvEQD1kt12}b, which shows that negative values of the nonlocal conductance (where the crossed Andreev reflections dominate over ballistic transfer) are realized in the strong interdot coupling limit solely when $\left|\varepsilon_{1}\right|<0.75$.

The energy levels (tunable by external gate potentials) along with the interdot coupling (transmitting electron pairing from QD$_2$ on QD$_1$) are thus controlling relationship between the single particle electron transfer and the anomalous electron-to-hole scattering processes.

\subsubsection{Thermopower}
\label{normal_thermopower}

We now study how different transport channels affect the thermoelectric properties. Usually in bulk materials a sign of the Seebeck coefficient reverses upon traversing from electron to hole dominated transport regions. The same behavior occurs in two-terminal metallic junctions with the single quantum dot, where the thermopower takes saw-tooth shape as a function of the energy level \cite{Weymann-2013,Manaparambil-2023}. In two-terminal nanostructures with quantum dot between a normal and  superconducting electrode, nonvanishing value of the Seebeck coefficient is obtained either at high temperatures or in presence of the Zeeman splitting \cite{Hwang-2016}. 

We can identify particle/hole dominant regions of our three-terminal setup by inspecting the local Seebeck coefficient and can distinguish between the ballistic and Andreev-type contributions, using the nonlocal coefficient. Fig.\ \ref{figSvEQD1kt12} shows variation of the local $S_{LL}$ and $S_{RL}$ Seebeck coefficients with respect to QD$_1$ energy level obtained for several values of $t_{12}$. In analogy to two-terminal junctions, the local Seebeck coefficient has negative sign for $\varepsilon_{1}<0$ and is positive for $\varepsilon_{1}>0$. The interdot coupling, $t_{12}$, causes only a flattening of its saw-tooth shape. 

The nonlocal Seebeck coefficient, on the other hand, is very sensitive on $t_{12}$. For small $t_{12}$ the nonlocal $S_{RL}$ has a shape typical for a metallic dot, i.e. $S_{RL}>0$ for $\varepsilon_{1}<0$ and $S_{RL}<0$ for $\varepsilon_{1}>0$ (see e.g. \cite{Weymann-2013,Wojcik-2014}). For the strong interdot coupling limit (when the molecular Andreev bound states are formed on QD$_1$ and QD$_2$) we observe  divergence and sign reversal of $S_{RL}$. This behavior is typical for the superconducting-proximity-effect  regime, where $S_{RL}<0$ for $\varepsilon_{1}<0$ and $S_{RL}>0$ for $\varepsilon_{1}>0$, see Refs.\ \cite{Michalek-2016,Valentini-2015}. 

Such divergence point corresponds to changeover from the thermoelectricty dominated by the ballistic channel to the region dominated by the crossed Andreev reflections. This information is thus complementary to the one of local Seebeck coefficient.

\section{Setup with ferromagnetic leads}
\label{sec:polarLeads}

\begin{figure}
\centering
\includegraphics[width=1\linewidth]{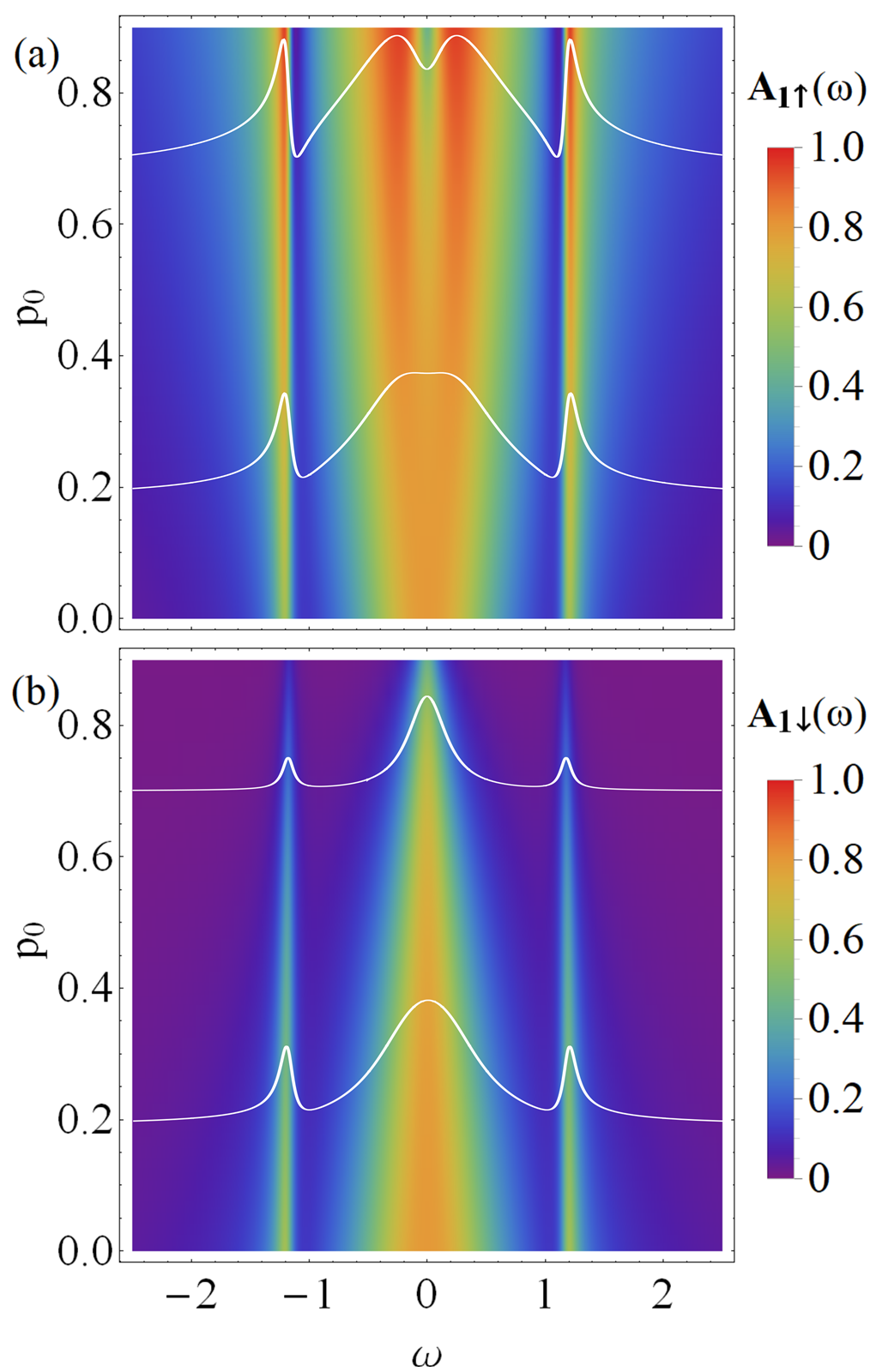}
\caption{
The normalized spectral function $A_{1\sigma}(\omega)$ of QD$_1$ as a function of the polarization parameter $p_0$. Numerical results are obtained for  $\varepsilon_{1}=\varepsilon_{2}=0$, $\Gamma_{S}=2$, $\Gamma_{L}=\Gamma_{R}=0.5$, $t_{12}=0.5$ and $T=0$. White lines show the profile of the spectral function of the weakly ($p_0=0.2$) and strongly polarized ($p_0=0.7$) system, respectively.} 
\label{figA1polar}
\end{figure}

We now consider the influence of electrode polarization, $p_0$, on the properties of our system. Magnetically polarized electrodes can be expected to induce spin-dependent features, analogous to those due to the Zeeman splitting. Since magnetism and superconductivity compete with each other, its hence important to analyze how they show up in the quasiparticle spectra of QD$_{1}$ and  how they affect the transport properties of our superconducting nanostructure. 

\subsubsection{Polarized spectrum}
\label{sec:polarized_spectrum}

To proceed, let us analyze influence of the polarization on quasiparticle spectra of the quantum dots, focusing on QD$_{1}$ because of its crucial role for the transport properties.
Fig. \ref{figA1polar} shows variation of the spin-resolved spectrum of QD$_1$ with respect to $p_0$ obtained for $\varepsilon_1=0=\varepsilon_2$ and the interdot coupling $t_{12}=0.5$, where we can clearly resolve the quasiparticle states of odd/even parity. For $p_0$=0 the spectrum consists of the central peak at $\varepsilon_1$ and two  Fano-type resonances near $\pm\Gamma_S/2$ driven by the superconducting proximity effect. Upon increasing the polarization the spectral function of spin-$\uparrow$ electrons absorbs more and more spectral weight and simultaneously the central peak gradually splits. In contrast, the outer resonances do not change their positions, proving that they correspond to the spinless BCS-type states $u^2 \left| 0\right> \pm v^2 \left| \uparrow\downarrow\right>$. As far as the spectrum of spin-$\downarrow$ electrons is concerned, the polarization of external leads depletes its spectral weight. 
Both these phenomena have strong effect on the transport properties, especially on their spin-sensitive versions.

\subsubsection{Conductance of polarized system}

\begin{figure}[tb] 
\centering
\includegraphics[width=1\linewidth]{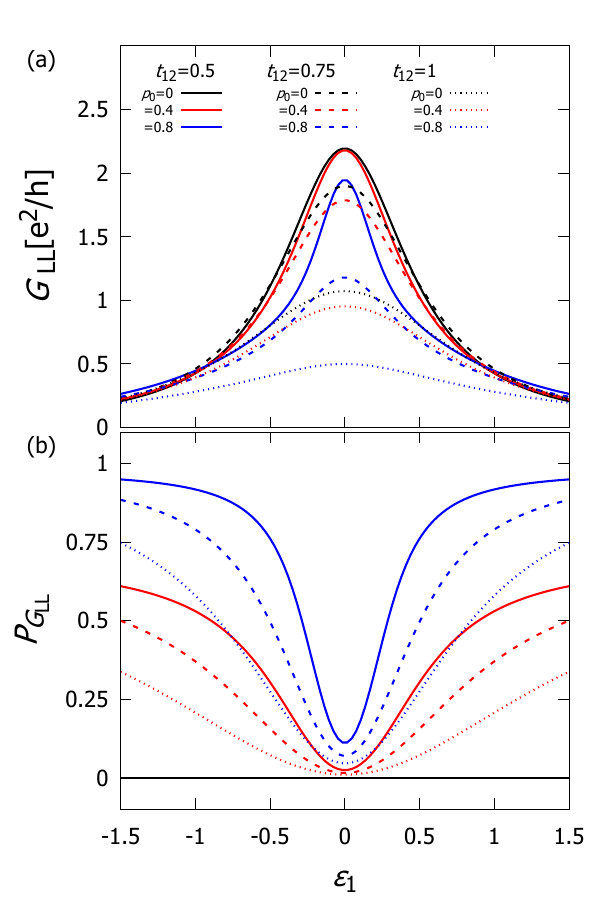}
\caption{
The local conductance $G_{LL}$ (a) and its polarization $P_{G_{LL}}$ (b) as function of the energy level $\varepsilon_1$ obtained for various $t_{12}$ and $p_0$, as indicated. Results are obtained for: $\varepsilon_{2}=0$, $\Gamma_{S}=2$, $\Gamma_{L}=\Gamma_{R}=0.5$, and $k_BT=0.05$.} 
\label{figGLLved1polar}
\end{figure}

Polarization affects the charge transport, both in the ballistic channel and the Andreev-type processes. In the first case its influence comes predominantly via renormalization of the low-energy spectral functions $A_{1\sigma}(\omega =0)$, whereas the electron-to-hole scatterings are sensitive to the modified electron pairing on QD$_{1}$. Evolution of the quasiparticle spectrum (discussed in Sec.\ \ref{sec:polarized_spectrum}) suggests thus detrimental influence of the polarization on the differential conductance.

Figure \ref{figGLLved1polar} shows variation of the local conductance $G_{LL}$ (a) and its polarization $P_{G_{LL}}$ (b) with respect to the energy level $\varepsilon_{1}$ for several $t_{12}$ and $p_0$. As expected, for all values of the interdot coupling, the local conductance is suppressed by $p_0$. The polarized conductance, Fig.\ \ref{figGLLved1polar}b, proves that the spin-resolved ballistic channel is mostly affected when $\varepsilon_1$ is distant from the symmetry point. The Andreev channels (simultaneously involving both of the spin components) are responsible for suppressing the local conductance when $\varepsilon_1 \sim 0$.

\begin{figure}[tb] 
\includegraphics[width=1\linewidth]{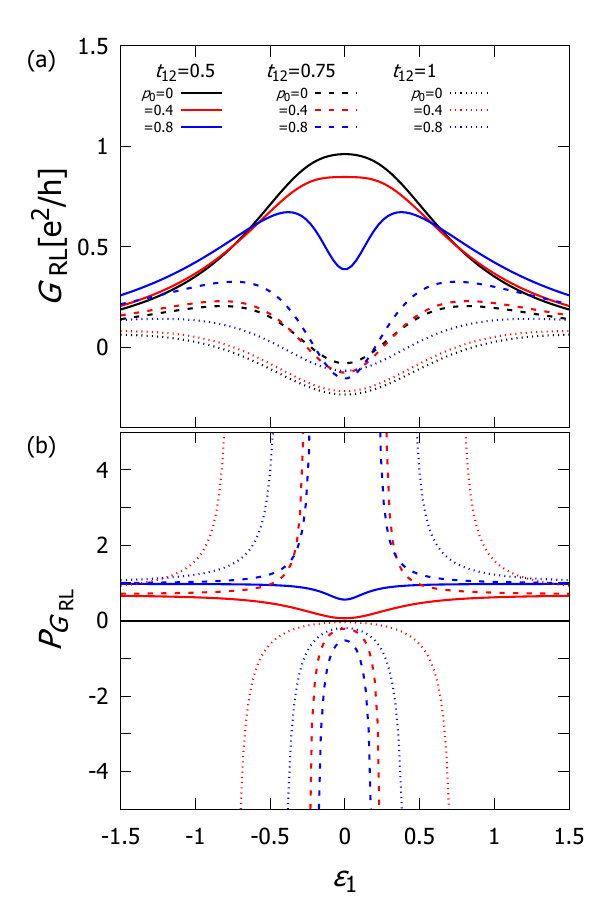}
\caption{Variation of the nonlocal conductance $G_{RL}$ (a) and its polarization $P_{G_{RL}}$ (b) with respect to the energy level $\varepsilon_1$ obtained for several values of $t_{12}$ and $p_0$, using the model parameters: $\varepsilon_{2}=0$, $\Gamma_{S}=2$, $\Gamma_{L}=\Gamma_{R}=0.5$, and $k_BT=0.05$.}
\label{figGRLved1polar}
\end{figure}

Figure \ref{figGRLved1polar}a presents the nonlocal conductance $G_{RL}$ obtained for different values of $t_{12}$ and $p_0$. Again we observe, that for small $\left|\varepsilon_1\right|$ the polarization $p_0$ suppresses mainly the crossed Andreev reflections. 
The polarized nonlocal conductance, Fig.\ \ref{figGRLved1polar}b, reveals divergence points which indicate changeover of the dominant transport channel. Such points are sensitive to the polarization, and occur only for sufficiently strong interdot coupling $t_{12}>0.5$.
%
The differential conductance together with its polarized version provide valuable information about the leading transport channel of our multi-terminal superconducting junction, what is particularly useful in presence of the external fields.

\begin{figure}
\centering
\includegraphics[width=1\linewidth]{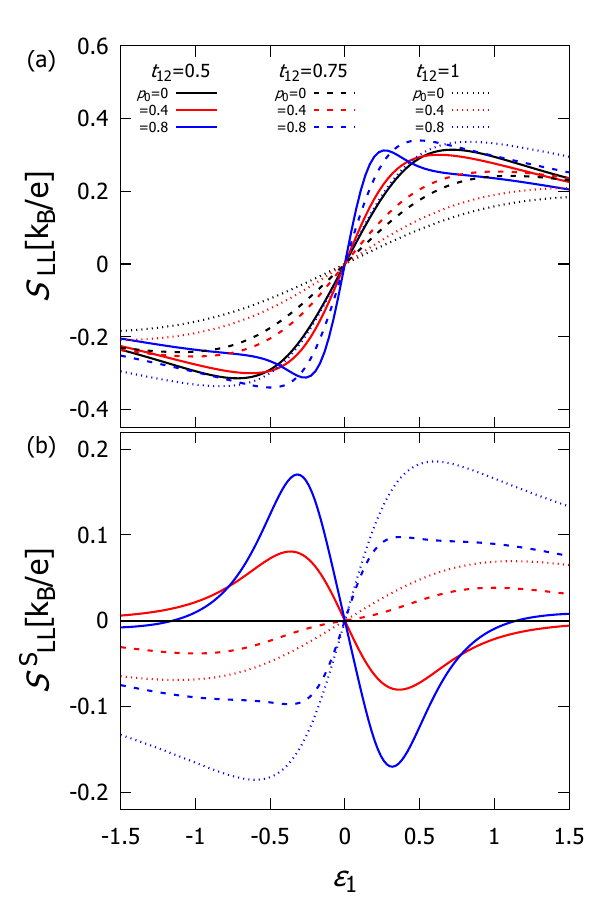}
\caption{
The local $S_{LL}$ (a) and the spin-resolved thermopower $S^S_{LL}$ (b) as a function of the QD$_1$ energy level obtained for several values of $t_{12}$ and $p_0$, assuming the model parameters: $\varepsilon_{2}=0$, $\Gamma_{S}=2$, $\Gamma_{L}=\Gamma_{R}=0.5$, and $k_BT=0.05$.}
\label{figSLLved1polar}
\end{figure}

\subsubsection{Thermopower  of polarized system}
\label{polarized_thermopower}

Polarization of the external leads has an influence on the Seebeck coefficients. Let us recall, that in N-QD-N junctions the thermopower is determined by a slope of the quantum dot spectral function $\partial A_\sigma(\omega)/\partial \omega$ at the Fermi level ($\omega=0)$ \cite{Costi-2010}. In the weak interdot coupling limit of our setup and in absence of polarization, the spectrum of QD$_1$ is represented by a Lorentzian peak centered at $\varepsilon_1$ (Fig.\ \ref{figA1kt12}) therefore the local thermopower (Fig.\ \ref{figSLLved1polar}a) has this usual saw-tooth shape.  Polarization induces splitting of the quasiparticle state at $\omega \sim \varepsilon_{1}$ (Fig.\ \ref{figA1polar}), what partly flattens the net thermopower.

Qualitative changes, however, can be noticed in the spin-resolved Seebeck coefficient defined in Eqn.\ (\ref{eq:SS}). In presence of polarization, for $\varepsilon_1<0$ and the weak interdot coupling we obtain $\partial A_\up/\partial \omega(\varepsilon_F)>\partial A_\down/\partial \omega(\varepsilon_F)$, therefore the spin-resolved thermopower is positive. By increasing $t_{12}$, the polarization amplifies the splitting of $A_\up(\omega)$ sector and suppresses the spectral weight of  $A_\down(\omega)$. In effect, the thermopower $S_{LL}(S)$  (Fig. \ref{figSLLved1polar}b) reverses its sign due to a combined influence of the electron pairing and the polarization imposed on the central quantum dot.

The spin-resolved Seebeck coefficient would be thus a sensitive probe of the superconducting proximity effect (transmitted on QD$_1$ via interdot coupling) competing with the polarization effects.

\section{Summary}
\label{sec:summary}

We have studied the quasiparticle spectrum and transport properties of the double quantum dot embedded in T-shape geometry between two conducting (or ferromagnetic) leads and strongly hybridized with superconducting electrode. In this configuration the proximity effect induces the bound states on the quantum dot adjacent to superconductor (QD$_2$) which next is partly or entirely transmitted to the other dot (QD$_1$). In the weak interdot coupling regime such process leads to appearance of the Fano-like structures in the spectrum of QD$_1$, whereas the tightly bound quantum dots develop the molecular structure of their Andreev states. 
These phenomena qualitatively affect the charge transport and thermoelectric properties of the setup.

We have thoroughly analyzed influence of the interdot coupling,  $t_{12}$, on the transport coefficients which could be experimentally accessible. Specifically, we have examined variation of the local/nonlocal conductance and thermoelectric coefficients upon increasing $t_{12}$. We have found that in the weak coupling limit the ballistic transport becomes reduced (by destructive quantum interference) whereas the direct and crossed Andreev scatterings are gradually amplified (by the  superconducting proximity effect indirectly transmitted onto QD$_1$). This tendency is clearly reflected in the local electric conductance (Fig.\ \ref{figGLLGRL} and Fig.\ \ref{figGvEQD1kt12}) and is even more pronounced in the nonlocal conductance, revealing a competition of the ballistic electron tunneling with the crossed Andreev reflection (\ref{eq_18}). The latter one undergoes suppression upon increasing $t_{12}$, and eventually reverses its sign when the crossed Andreev reflections become the dominant transport channel.

We have also examined how the interdot coupling affects the local and nonlocal thermopower. For the weak interdot coupling we have found that the local Seebeck coefficient acquires the usual sawtooth shape as function of the quantum dot energy level (caused by changeover from the particle to hole dominated single particle charge transfer). In contrast, for the large coupling $t_{12}$, we have obtained qualitatively different behavior which manifests  the superconductivity-proximity-dominated regime. In particular, it reveals the divergence point.

In the setup with ferromagnetic electrodes the ballistic transport channel is dependent on individual spin components. For this reason we have investigated in detail the spin-resolved conductance and the Seebeck coefficients, both in their local and nonlocal versions. The resulting transport coefficient reveal subtle interplay between the spin-resolved ballistic transfer and the Andreev scatterings, which equally engage both spin components.
 
In summary, we have shown that charge-transport measurements would be able to probe the efficiency of the superconducting proximity effect imposed on the double quantum dot in the three-terminal hybrid structures. Transport properties could clearly distinguish the molecular bound states (when the interdot coupling is strong) from different situation, where the in-gap states are formed merely in one of the dots, while the second one is affected through interferometric (Fano-type) effects. These phenomena could be encountered also in other hybrid architectures, for instance using two quantum dots attached to the Majorana modes \cite{Calle-2020}. Furthermore, similar effects might show up when inspecting the parity of Andreev molecule of two quantum dots interconnected via a superconducting reservoir \cite{PhysRevResearch-2024,PRXQuantum-2025}. The latter scenario is nowadays considered as a possible mean for realization of the minimal Kitaev model, hosting the poor man's Majorana quasiparticles \cite{Kouwenhoven-2023}.

\begin{acknowledgments}
This research project has been supported by the National Science Centre (Poland) through the grant No.\ 2022/04/Y/ST3/00061.
\end{acknowledgments}


\section*{Data availability}
The datasets generated and analyzed during the current study are available from the repository \cite{data_2025}.

\appendix

\section{Correlation effects}
\label{sec:correlations}

\begin{figure}[b!] 
\centering
\includegraphics[width=1\linewidth]{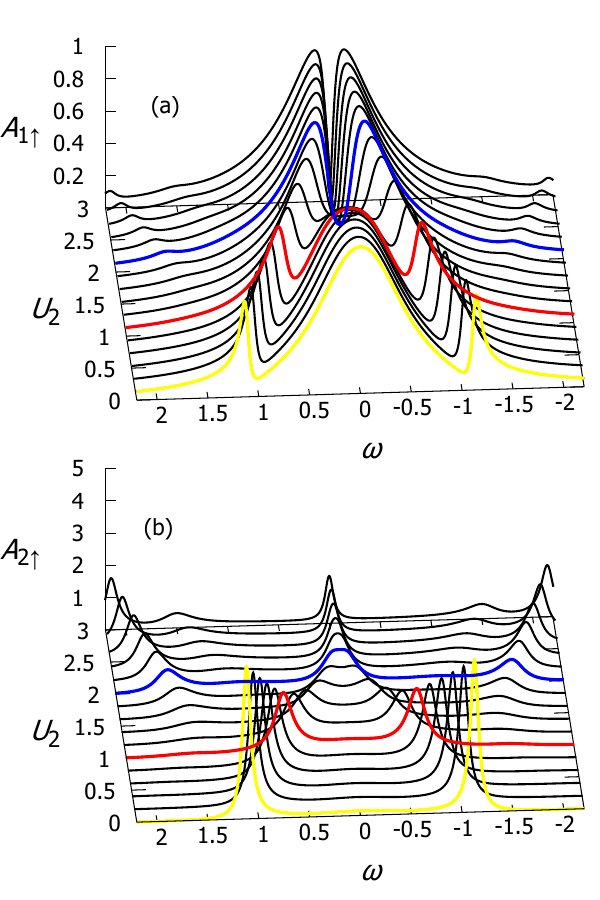}
\caption{
The normalized spectral function $A_{i\up}(\omega)$ of the first (a) and second (b) quantum dot as a function of the Coulomb repulsion $U_{2}$. Other parameters are: $\varepsilon_{1}=0$, $\varepsilon_{2}=-U_{2}/2$, $\Gamma_{S}=2$, $\Gamma_{L}=\Gamma_{R}=0.5$, $k_BT=0$, $p_0=0$, and $t_{12}=0.4$.} 
\label{figA1A2corr}
\end{figure}

Under specific conditions, the spectra of the quantum dots can additionally be influenced by Coulomb repulsion between opposite spin electrons $U_{i}d_{i\uparrow}^{\dagger}d_{i\uparrow}d_{i\downarrow}^{\dagger}d_{i\downarrow}$. We shall discuss the major effects arising from such interactions, inspecting their role at each dot ($i=1,2$) separately. 

The repulsive potential $U_2$ on the quantum dot directly attached to the superconductor can be expected to compete with the proximity effect. Signatures of this competition are evident already within the mean field approximation (valid for $U_2\ll \Gamma_S$), leading to renormalization of the on-dot pairing potential
%
$\frac{\Gamma_{S}}{2} \rightarrow \frac{\Gamma_{S}}{2}-U_{2} \left< d_{2\downarrow} d_{2\uparrow}\right>$.   
%
Using more sophisticated methods to treat the Hamiltonian of QD$_{2}$ \cite{Zonda-2023}, substantial changes of the Andreev bound states have been predicted. In particular, by varying $U_2/\Gamma_s$ the ratio or the energy level $\varepsilon_2$ the ground state of QD$_2$ eventually undergoes the quantum phase transition between the BCS-type (spinless) and the singly-occupied (spinfull) configurations \cite{Bauer-2007}. For the particle-hole symmetric case ($\varepsilon_2=-U_2/2$), this quantum phase transition occurs at $U_2=\Gamma_S$. In the strong interaction limit, $U_2 > \Gamma_S$, the on-dot pairing $\left< d_{2\downarrow} d_{2\uparrow}\right>$ is strongly (or completely) suppressed. The influence of such effects on the local and nonlocal conductance of the three-terminal junction comprising the single quantum dot has recently been addressed in Ref.\ \cite{Hashimoto-2024}. Here, in the setup with double quantum dot, the Coulomb potential could predominantly affect the transport channels contributed by the Andreev scattering. 

Figure \ref{figA1A2corr} shows the quasiparticle spectrum of spin-$\uparrow$ electrons of both quantum dots obtained for the particle-hole symmetric case, $\varepsilon_2=-U_2/2$ (neglecting the Coulomb potential $U_1=0$). Numerical results for the Green's function matrix have been obtained by the self-consistent second-order perturbation theory \cite{Vecino-2003,Yamada-2011,Gorski-2018}. Inspecting $A_{2\uparrow}(\omega)$ we can clearly notice the singlet-doublet changeover that occurs near $U_2\approx\Gamma_S$  (strictly speaking, it is no longer sharp due to the interdot coupling). In the spectrum of QD$_2$ adjacent to the superconducting lead (Fig.\ \ref{figA1A2corr} b), we observe two Andreev states which merge at $U_2 \rightarrow \Gamma_S$. In contrast, in the strong interaction limit, the Abrikosov-Suhl state develops at the Fermi level, $\omega=0$, originating from the antiferromagnetic exchange interactions of QD$_2$ with mobile electrons from external metallic leads \cite{Domanski-2016}.
In addition to this Kondo feature, there are also other quasiparticles with finite energies that represent molecular Andreev states \cite{Bauer-2007}. On the other hand, the spectrum of QD$_1$ (Fig.\ \ref{figA1A2corr}a) reveals completely different line shapes. At energies corresponding to the Andreev peaks (in the weak interaction limit, $U_2<\Gamma_s$) we observe resonant-like dip structures, and (in the strong interaction regime, $U_2>\Gamma_s$) another depleted region is observed around the Kondo state.

\begin{figure}[tb] 
\includegraphics[width=1\linewidth]{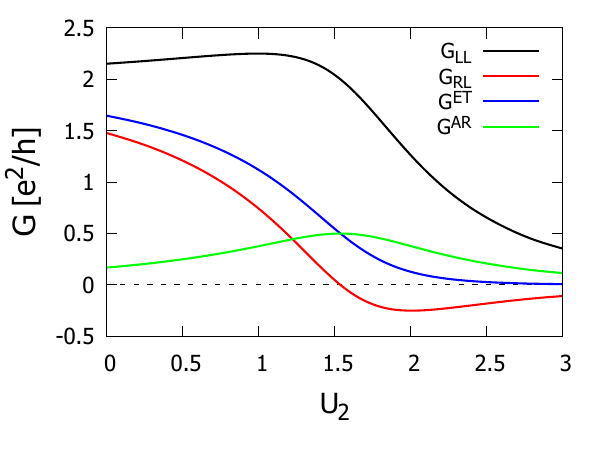}
\caption{
The zero-bias conductance (local $G_{LL}$ - black line, nonlocal $G_{RL}$ - red line, electron tunneling $G^{ET}$ - blue line and Andreev reflection $G^{AR}=G^{DAR}=G^{CAR}$ - green line) as a function of the Coulomb repulsion $U_{2}/\Gamma_{N}$. Results are obtained for: $\varepsilon_{1}=0$, $\varepsilon_{2}=-U_{2}/2$, $\Gamma_{S}=2$, $\Gamma_{L}=\Gamma_{R}=0.5$, $T=0$, $p_0=0$,
and $t_{12}=0.4$.}
\label{figGcorr}
\end{figure}

The influence of the Coulomb potential $U_2$ on the local and nonlocal transport properties of our system is presented in Fig.\ \ref{figGcorr}. We notice that upon traversing the singlet-doublet transition, the ballistic electron transfer (blue line in Fig.\ \ref{figGcorr}) becomes gradually suppressed. In contrast to this, the Andreev conductance achieves optimal values around this crossover region (where the molecular quasiparticle energies cross each other). Further consequence of this behavior is observed in the nonlocal conductance (displayed by a red line), which changes sign to negative values near the singlet-doublet crossover.

Concerning the Coulomb interactions $U_1$, its influence is qualitatively different from the one discussed above. Repulsive interactions at QD$_1$ mainly affect the ballistic transport channel, being less efficient for Andreev scattering. The resulting effect is hence reminiscent of the Coulomb blockade. Its signatures could be observed by enhancing the differential conductance at $\omega = \varepsilon_1$ and $\varepsilon_1+U_1$ due to charging effects. Furthermore, at low temperatures, the Abrikosov-Suhl peak could be formed induced by the effective exchange interactions between the mobile electrons of the metallic leads and the localized electron on QD$_1$. Such Kondo physics would lead to enhancement of the zero-bias conductance of the single electron (ballistic) channel. Further indirect effects on the proximity-induced electron pairing would be less meaningful unless the interdot coupling is strong. Major aspects related to the latter mechanism have been studied by Orellana {\it et al.} \cite{Calle-2020}, therefore we skip their discussion. 

\section{Polarized transport coefficients}
\label{sec:coefficients}

\begin{figure}[tb] 
\centering
\includegraphics[width=0.95\linewidth]{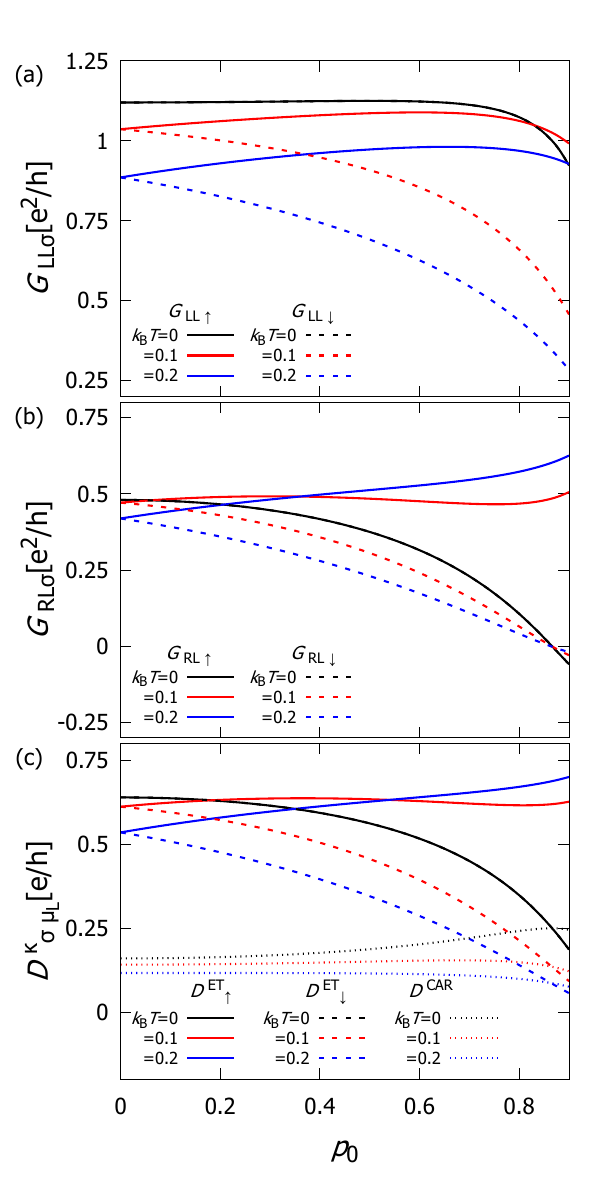}
\caption{
Variation of the local $G_{LL}$ (a) and nonlocal $G_{RL}$ (b) spin-dependent  conductivity and the transport coefficients $D_{\sigma\mu_L}^{\kappa}$ (c) versus the polarization, $p_0$, obtained for several temperatures using the model parameters: $\varepsilon_{1}=\varepsilon_{2}=0$, $\Gamma_{S}=2$, $\Gamma_{L}=\Gamma_{R}=0.5$ and $t_{12}=0.5$.} 
\label{figGLRvp0}
\end{figure}

In this part we briefly consider influence of temperature on the transport coefficients, originating from the Fermi-Dirac distribution function entering the expressions for charge current.
Fig.\ \ref{figGLRvp0} shows the local $G_{LL}$ and nonlocal $G_{RL}$ spin-dependent conductivity with respect to the polarization of electrodes $p_0$ obtained in the electron-hole symmetric case, $\varepsilon_{1}=0=\varepsilon_2$, and for several temperatures. For $T=0$ (black lines), the zero-bias conductivity is proportional to the transmittance at $\omega=0$, therefore  $G_{LL(RL)\up}=G_{LL(RL)\down}$. 
By increasing the polarization, we observe suppression of $T^{ET}_{\sigma}(0)$ and slight enhancement of $T^{DAR(CAR)}_{\sigma}(0)$. Consequently, the local conductance $G_{LL\sigma}$ weakly depends on $p_0$, whereas the nonlocal conductance $G_{RL\sigma}$ reveals gradual reduction. At finite temperatures, $T>0$, we obtain spin-dependent conductance $G_{LL(RL)\up} \neq G_{LL(RL)\down}$ for arbitrary values of $p_0\neq 0$. In such a case, nonvanishing spin polarization of the local and nonlocal conductance can be observed.

\bibliography{myBib}

\end{document}